\documentclass[conference]{IEEEtran}
\usepackage{eso-pic}
\usepackage{hyperref}
\usepackage{xcolor}

\hypersetup{
    colorlinks=true,
    urlcolor=blue,
}

\AddToShipoutPictureFG*{
  \put(0,60){ 
    \makebox[\paperwidth]{ 
      \centering
      \footnotesize The code for this paper is available at: 
      \href{https://github.com/XXX}{\texttt{https://github.com/aymenhamrouni/hybrid-rf-woc-aoi-optimization}}
    }
  }
}

\makeatletter
\def\ps@IEEEtitlepagestyle{%
  \def\@oddfoot{\mycopyrightnotice}%
  \def\@evenfoot{}%
}
\def\mycopyrightnotice{%
  {\footnotesize KU LEUVEN ~\copyright~2024\hfill
  }
  \gdef\mycopyrightnotice{}
}
\usepackage[utf8]{inputenc}

\usepackage{optidef}
\usepackage{blindtext}
\usepackage{eso-pic}
\IEEEoverridecommandlockouts
\usepackage{cite}
\usepackage{amsmath,amssymb,amsfonts}
\usepackage{algorithmic}
\usepackage{graphicx}
\usepackage{comment}
\usepackage{textcomp}
\usepackage{xcolor}
\usepackage{caption}
\usepackage{subcaption}
\def\BibTeX{{\rm B\kern-.05em{\sc i\kern-.025em b}\kern-.08em
    T\kern-.1667em\lower.7ex\hbox{E}\kern-.125emX}}
\usepackage{eso-pic}

\usepackage{geometry}
\usepackage[font=footnotesize]{subcaption}
\usepackage[font=footnotesize]{caption}
\usepackage{flushend}

\begin{document}
\title{ \huge{AoI in Context-Aware Hybrid Radio-Optical IoT Networks}}

\author{\IEEEauthorblockN{Aymen Hamrouni, Sofie Pollin, and Hazem Sallouha}
\IEEEauthorblockA{\small 
Department of Electrical Engineering (ESAT) - WaveCoRE, KU Leuven, Leuven, Belgium\\
Email: \{aymen.hamrouni,sofie.pollin,hazem.sallouha\}@kuleuven.be} }

\maketitle

\begin{abstract}
With the surge in IoT devices ranging from wearables to smart homes, prompt transmission is crucial. The Age of Information (AoI) emerges as a critical metric in this context, representing the freshness of the information transmitted across the network.  This paper studies hybrid IoT networks that employ Optical Communication (OC) as a reinforcement medium to Radio Frequency (RF).  We formulate a non-linear convex optimization that adopts a multi-objective optimization strategy to dynamically schedule the communication between devices and select their corresponding communication technology, aiming to balance the maximization of network throughput with the minimization of energy usage and the frequency of switching between technologies. To mitigate the impact of dominant sub-objectives and their scale disparity, the designed approach employs a regularization method that approximates adequate sub-objective scaling weights. Simulation results show that the OC supplementary integration alongside RF enhances the network's overall performances and significantly reduces the Mean AoI and Peak AoI, allowing the collection of the freshest possible data using the best available communication technology.

\end{abstract}

\begin{IEEEkeywords}
IoT, Hybrid RF-OC, AoI, Optimization
\end{IEEEkeywords}

\section{Introduction}
In recent years, the proliferation of Internet of Things (IoT) devices has revolutionized the way we live and work. From wearable gadgets that monitor our health to smart home systems that enhance our living environment, the ubiquity of IoT technologies is undeniable~\cite{10433153}. A critical aspect of IoT applications is the collection and transmission of data on time. In a healthcare setting, for example, IoT enables remote monitoring systems to track patients' vitals in real-time. These systems continuously gather data on heart rate, blood pressure, and glucose levels, which must be sent promptly to enable accurate medical interventions and personalized care plans. One of the paramount Key Performance Indicators (KPIs) in IoT ecosystems is the Age of Information (AoI), which measures the freshness of the data being transmitted to and processed by a receiver. AoI is defined as the difference between the current time and the time at which the last successfully received update was generated at the source~\cite{10032496,10000990}. This metric is important, especially when the data is time-sensitive and the network is energy-limited. Finding a balance between keeping the information as fresh as possible and minimalist energy consumption is crucial for a smooth IoT network operation.

Managing this information freshness versus energy trade-off in a single-technology IoT network presents a significant challenge. There have been some adaptive strategies that adjust the frequency of updates based on the criticality of information and current network conditions~\cite{10049773}. However, single-technology IoT networks, which predominantly rely on Radio Frequency (RF) communication are unscalable due to the limited RF resources. This scalability problem leads to an unpredictable effect on both AoI and energy efficiency. The integration of a secondary technology into the IoT network, such as Optical Communication (OC), can be a promising solution to such challenges~\cite{katz2024towards}. OC technology can be designed as a reinforcement communication medium to the existing RF mediums to kick in and accelerate data transmission when critical updates are necessary. This enables IoT networks to achieve more flexible and efficient performance by leveraging: 1) the high-speed data transmission capabilities of OC and 2) the energy efficiency and physical barrier penetration of RF~\cite{2}. Indeed, most IoT devices are currently equipped with RF communication, but an OC extension is inexpensive and easy to set up~\cite{10008721}. The integration of OC serves as a supplementary rather than a primary communication technology, promoting a harmonious balance that optimizes energy consumption, ensures data freshness, and accommodates the dynamic conditions of IoT environments\cite{katz2024towards,9912428}. This being said, minimizing the AoI metric in an optimization problem with such hybrid settings represents challenge in itself. This is mainly due to the AoI's inherently non-linear~\cite{9380899} and non-convex nature concerning decision variables such as scheduling, packet generation rates, and resource allocation~\cite{8883204}. Such characteristics render the AoI tractable only in highly simplified settings.

Significant research has been conducted on hybrid RF-OC systems. Fakirah et al.~\cite{1} explored a visible light with an RF system that uses a set of hybrid access points to exchange information between the vehicles traversing the roundabout and roadside units in the vehicle-to-infrastructure mode. Obeed et al.\cite{3} designed an iterative optimization-based algorithm for load balancing and power allocation schemes for a hybrid light and RF system consisting of one RF access point and multiple OC access points. Xiao et al.\cite{5} investigated a hybrid downlink system that simultaneously uses visible light communication and RF with a cognitive-based resource allocation policy.
On the other hand, studies focusing on the AoI include Lui et al.~\cite{9755889}, where the average AoI in a wireless-powered MEC system is assessed. 
Chen et al.~\cite{10032496} studied the AoI in a multi-channel network-sided information. Xu et al.\cite{9239322} analyzed the Peak AoI across two scenarios: one considering a buffer size of one and infinite, and the other examining the infinite buffer size case. To the best of our knowledge, no existing work has investigated the impact of hybrid RF-OC devices on the AoI in IoT networks.


In this paper, we formulate a non-linear convex optimization algorithm within a hybrid RF-OC network that seeks to 1) maximize the overall throughput across the network, 2) minimize the energy consumption, and 3) minimize the frequency of technology switch between RF and OC within the network. A unique feature of this approach is the dynamic selection mechanism that enables nodes to choose the most appropriate technology—RF or OC—at any given moment. To address the multifaceted nature of this multi-objective system, we conduct a regularization analysis to select the weights of the individual sub-objectives and avoid dominance or saturation. While the AoI was not minimized directly in the optimization problem, simulation results show that the RF-OC hybrid system achieves lower Mean AoI (M-AoI) and Peak AoI (P-AoI) compared to a single-technology RF system. Moreover, our problem formulation with adequate regularization inherently leads to improved information freshness in the RF-OC setup while leveraging conventional convex optimization.

\section{System Model}
In this section, we introduce the system model and present the overall network and communication setups. 

\subsection{Network Setup}
\begin{figure}[t]
\begin{center}
\includegraphics[width=0.4\textwidth]{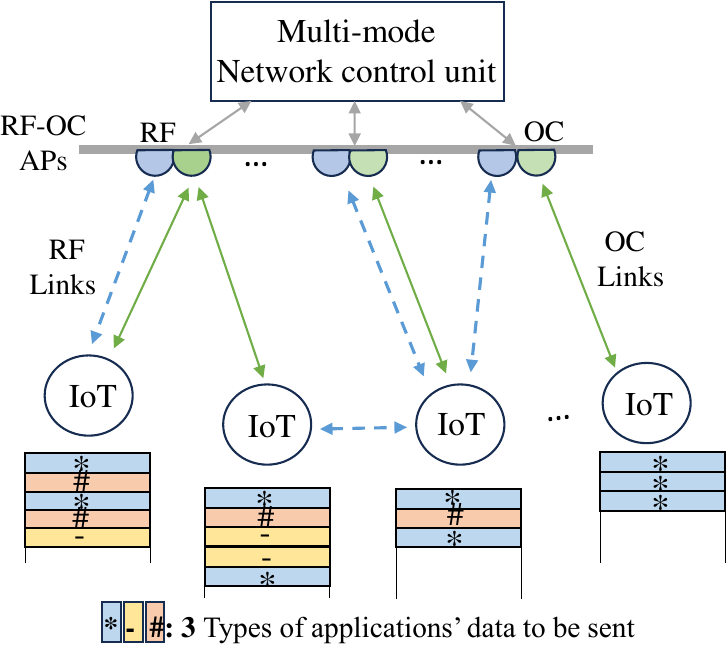}
\end{center}
\caption{System Model. This example shows the network using RF-OC to exchange 3 different types of applications' data. }
\label{fig0}
\end{figure}

We consider an IoT network comprising a set of IoT nodes and a set of Access Points (APs) having OC and RF capabilities. Nodes communicate with APs using either OC or RF, whereas only RF is used for inter-node communication. The system model with an RF-OC hybrid configuration is illustrated in Fig.~\ref{fig0} with different IoT nodes transmitting multiple types of application's data.  Let $N_d$ be the number of IoT nodes and $N_{APs}$ be the number of APs in the system. The total number of devices in the network is denoted by $N=N_d+N_{APs}$. Let $L$ be the number of different types of data in the system. For example, in the case of healthcare settings, IoT devices can exchange two types of messages, say heart rates and oxygen levels. In this case, $L$ is set to $2$. Let $M$ be the set of possible communication technologies. In our system, $M=\{0,1\}$ with $m=0, m \in M$ representing RF, and $m=1, m \in M$ denoting OC. Let $T$ be a discrete time interval representing a time span split into equal-length discrete time intervals. The system is modeled over a discrete time horizon, segmented into time steps $\forall k \in \{1, \ldots, T\}$. Each IoT device $i \in N$ possesses energy $E^m_i$ for technology $m$. Indeed, we suppose that each device can allocate separate energy for each communication technology~\footnote{This assumption leads to a more scalable system where IoT devices can be seen as two joint devices, each providing a communication technology.}. Let ${E_r}^m$ denote the energy consumed by any device to receive a message using technology $m$ and  ${E_s}^m$ be the energy consumed by any device to send a message using technology $m$. We suppose that devices have equal energy consumption when sending or receiving using either communication technology. 

\subsection{Communication Setup}

We consider a 3-D visibility matrix $V_{m}, m \in M$ where the elements of this matrix are $v^m_{i,j,k}$ representing the probability of successful communication between any device $i$ and device $j$ at time step $k$. This visibility matrix captures the dynamic communication behavior that changes over time $k$. For example, the communication channel for RF can change with time, and hence, $v^0_{i,j,k}$ for each pair $i$ and  $j$ also dynamically change with a time step $k$.  Also, $V_{m}, m \in M$ controls the communication flow between any two devices in the network. For example, as IoT nodes cannot communicate with each other using OC, this can be controlled by setting $v^1_{i,j,k}=0, \forall (i,j) \in N_{d}^2, \forall k \in T$ between any IoT nodes. Similarly, and as all communications between APs are not considered in our paper, $v^m_{i,j,k}=0,  \forall (i,j) \in N_{APs}^2, \forall m \in M, \forall k \in T$. The matrix $V_{m}, m \in M$ is symmetric and could be modeled as a statistical matrix based on the history of previous packet loss rates, for example. We also consider another 3-D communication matrix $P$ where the elements of this matrix are $\rho^k_{i,j}$ indicating if device $i$ wants to send a message to device $j$ at time $k$. We denote $\Phi_{i,j}=\{\phi^{i,j}_{1,l},\phi^{i,j}_{2,l},\cdots,\phi^{i,j}_{|\Phi_{i,j}|,l}\}$ the set of messages to be sent from device $i$ to device $j$. The term ${\phi^{i,j}_{1,l}}$, for example, represents the set of time steps of message $1$ with type $l$, that needs to be sent from device $i$ to device $j$. Each message in $\Phi_{i,j}$ between any two devices in the network must be sent at most one time and has a specific time length representing its generation time and discard time. The discard time represents the maximum time at which the message could be sent\footnote{A message is considered old when $k$ reaches its discard bound with no successful sending. This discard time is an upper bound for how long the message can remain in the queue without being sent.}. We suppose that there is no time overlap between any two distinct messages for a given device. In other words, $\phi^{i,j}_{a,l} \cap \phi^{i,j}_{b,l}=\emptyset, \forall a \neq b$, and $i,j,l$.  Also, we assume that each generated message can have a single type of data. Hence. $\phi^{i,j}_{f,l} \cap \phi^{i,j}_{f,l'}= \emptyset, \forall l \neq l'$. Let's denote $\phi^{i,j}_{{a,l}^s}$ as the starting time step and $\phi^{i,j}_{{a,l}^e}$  as for the ending time step for message $\phi^{i,j}_{a,l}$, respectively. As an example, and for $\phi^{0,2}_{1,3}=\{3,4,5\}$, device $0$ can send its first message, which is with type $3$, to device $2$ at any of the following time steps: $\phi^{i,j}_{{1,3}^s}=3$, $k=4$ or $\phi^{i,j}_{{1,3}^e}=5$. If the message is not sent before $k=5$, it is considered deprecated and will be discarded. 

\section{Proposed RF-OC Optimization}
This section defines the optimization problem in RF-OC IoT settings by outlining the decision variables, the constraints, and the overall objective function.

\subsection{Decision Variables}

In order to indicate the chosen 2-tuples (sender, receiver) and to keep track of assigned communication technology and the communication time step, we introduce a main binary decision variable $x_{i,j,m,k}$ defined as follows:

\begin{align} \label{x}
& x_{i,j,m,k}= 
     \begin{cases}
       \text{1,} &\,\text{if device $i$ transmit a message to device $j$} \\  & \text{using technology $m$ at time step $k$,} \\
       \text{0,} &\,\text{otherwise,} \\ 
     \end{cases}\notag\\
 &\hspace{2cm}\forall \, (i,j) \in N, \forall \, m \in M, \forall k \in T.
     \end{align}

To keep track of which technology a device $i$ is using at time step $k$, we introduce an \textit{endogenous} binary variable $s_{i,k}$, that is 0 if device $i$ using RF, and 1 if it is using OC. In the event of a device not communicating at all at time step $k$, the value $s_{i,k}$ is the same as $s_{i,k-1}$. Initially, every device starts with $s_{i,0}=0$. To get the relative time position within a message's valid window when the message is sent from device $i$ to device $j$, we introduce another \textit{endogenous} decision variable $\delta_{i,j,k}$, which has a strictly non-negative integer value. This decision variable is directly lined to starting time  $\phi^{i,j}_{{a,l}^s}$ and ending time $\phi^{i,j}_{{a,l}^e}$ of any two messages exchanged between devices $i$ and $j$.
Throughout the paper, we assume the absence of the
propagation delay, queuing delay, and processing delay and consider only the transmission delay as a single time step unit. For example, if device $i$ communicated with device $j$ at time step $k=\phi^{i,j}_{{a,l}^s}$ when the message has just been generated, $\delta_{i,j,k}$ is $1$. In case the message becomes stale and there has been no communication, the value of $\delta_{i,j,k}$  is set to be a fixed integer that is strictly greater than the largest message in the overall system (i.e. max($\|\phi^{i,j}\|$)).


Among the various metrics to quantify AoI\cite{10000990}, 
the linear AoI stands out as the most widely used and recognized due to its straightforwardness and comprehensive insights. If the freshest update from a device $i \in N$ at time $k \in T$ is generated at time $u(k) = \phi^{i,j}_{{f,l}^{s}}$, the AoI at the receiver $j \in N$ is defined as $\Delta(k) = k-u(k)$, which is the time elapsed since the generation of the last received update. The AoI linearly increases at a unit rate with respect to $k$, except some reset jumps to a lower value at points when the receiver receives a fresher update from the sender. In our defined settings, the AoI can be extrapolated from $\delta_{i,j,k}$ and $y_{i,j,f}$ to calculate the difference between the system's current time and the generation time of the last successful received message for a specific type $l$.

\subsection{Constraints}
To ensure that each IoT device within the network can communicate with at most another device and use either RF or OC at each time step, we include the following constraints:
\begin{equation}
\begin{split}
    \sum_{j=1}^{N} \sum_{m \in M} x_{i,j,m,k} \leq 1, \quad \forall i , \forall k 
       \\
        \sum_{i=1}^{N}\sum_{m \in M} x_{i,j,m,k} \leq 1, \quad \forall j , \forall k 
    \end{split}   
    \label{const1}
    \end{equation}
To enforce that communication between devices can happen only when there is an actual message to be transmitted, we add the following constraint:
\begin{equation}
\begin{split}
    \rho^k_{i,j} \geq x_{i,j,m,k}, \quad \forall i, j, \forall m,
    \forall k 
    \end{split}
 \label{const2} \end{equation}
The following constraint ensures that if a device opts to communicate using a communication technology, the communication must be feasible. We set a pre-defined threshold $\sigma_{m}, m \in M$ that determines the minimum value that must be set by the visibility matrix $V_{m}, m \in M$ for the communication to be considered feasible.
\begin{equation}
\begin{split}
   v^m_{i,j,k} \geq \sigma_{m} \cdot x_{i,j,m,k}, \quad \forall i, j , \forall m, \forall k
 \end{split}
  \label{const3} \end{equation}
The following constraint prohibits devices from communicating with themselves. Despite the inherent assumption that can be made in the communication matrix $P$ that denies self-communication, the following constraint explicitly enforces that rule within the optimization model.
\begin{equation}
\begin{split}
    x_{i,i,m,k} = 0, \quad \forall i , \forall m, 
 \forall k 
 \end{split}
  \label{const4} \end{equation}
Any IoT device $j$ engaged in communicating at a given time step with another device $i$ cannot simultaneously communicate with another device $j'$. This is set by the following constraint:
\begin{equation}
\begin{split}
    x_{i,j,m,k} + x_{j,j',m',k} \leq 1, \quad \forall (i, j, j') \in \{1, \ldots, N\}^3, \\ \forall (m, m') \in  M^2, \forall k \in \{1, \ldots, T\}
 \end{split} \label{const5} \end{equation}
The following energy consumption constraints prevent devices from exceeding their energy budget allocated to their communication technology.
    \begin{equation}
    \begin{split}
        \sum_{j=1}^{N} \sum_{k=1}^{T} x_{i,j,m,k} \cdot \left ( E_s^m +  E_r^m \right ) \leq E^m_i, \quad  \forall i  , \forall m 
    \end{split}  \label{const6} \end{equation}
Any device in the system can send multiple messages to a single receiver, and each of these messages can have different types of data. As each message has a duration of validity over several time steps, as discussed in the previous section, we impose the following constraints so that a message, regardless of its data type, can be sent at most once within its valid time window:
    \begin{equation}
    \begin{split}
        \sum_{m=1}^{M} \sum_{k \in \phi^{i,j}_{f,l}} x_{i,j,m,k} \leq 1, \quad \forall i, j, \forall f, \forall l
    \end{split} 
     \label{const7} \end{equation}
where $\phi^{i,j}_{f,l}$ represents the valid time slots of message $f$ with data type $l$ that should be sent from device $i$ to device $j$.

The following two constraints extract the time elapsed between the generation time of the message from a device $i$ and its sending time step. If the message was sent successfully, this duration is $(k -  \phi^{i,j}_{{u,l}^s})+1$ between the current time and the generation time step of message $u$ and counting the unit transmission delay. This constraint is written as follows: 
\begin{equation}
\begin{split}
&\quad \sum_{m=1}^M x_{i,j,m,k} = 1 \quad \rightarrow \quad \delta_{i,j,k} = (k -  \phi^{i,j}_{{u,l}^s})+1 , \\ & \forall k \in \phi^{i,j}_{u,l},  \forall l, \forall u \in \{1, \ldots, |\Phi_{i,j}|\}, \forall i,j) 
 \end{split}
  \label{const8} \end{equation}
with $\rightarrow$ representing a conditional statement. In the absence of transmission, the delay $\delta_{i,j,k}$ is set to an upper bound $\tau$ that represents a strict bound on the time duration of the longest message in the system.
\begin{equation}
\begin{split}
& \quad \sum_{m=1}^M x_{i,j,m,k} = 0 \quad  \rightarrow \quad \delta_{i,j,k} = \tau , \\  &\forall k \in \phi^{i,j}_{u,l}, \forall l,  \forall u \in \{1, \ldots, |\Phi_{i,j}|\},  \quad \forall i,j 
 \end{split}
  \label{const9} \end{equation}
To keep track of the communication technology switch, we add the following constraint:
\begin{equation}
\begin{split}
&\sum_{i=1}^{N} (x_{i,j,m,k} + x_{j,i,m,k}) = 1  \rightarrow s_{j,k} = m, \quad  \forall j , \forall m, 
\forall k 
 \end{split}
  \label{const10} \end{equation}
The value of $s_{j,k}$ must be the same as $s_{j,k-1}$ in the absence of transmission at time step $k$ using technology $m$. This is included in the following two equations:
\begin{equation}
\begin{split}
&\sum_{m=1}^{M} \sum_{i=1}^{N} (x_{i,j,m,k} + x_{j,i,m,k}) = 0 \land s_{j,k-1} = 0  \rightarrow s_{j,k} = 0, \\
&\sum_{m=1}^{M} \sum_{i=1}^{N} (x_{i,j,m,k} + x_{j,i,m,k}) = 0 \land s_{j,k} = 1   \rightarrow s_{j,k-1} = 1, \quad \\ & \forall j , \forall m, 
\forall k 
 \end{split}
  \label{const11}
  \end{equation}


with $\land$ being the logical AND. Indeed, the previous five constraints, from \eqref{const8}  to \eqref{const11}, are inherently non-linear, but we can approximate and enforce these conditions in linear programming by using the Big-M method~\cite{articless}. The Big-M method manages conditional constraints by introducing artificial variables and a large penalty factor to the objective function. This penalty ensures that the artificial variables, used to initially find a feasible solution, are driven out of the final solution.

\subsection{Problem Formulation}
Given a hybrid RF-OC IoT network with devices engaged in exchanging messages over discrete time steps, the objective function is designed to minimize the overall energy consumption in the network with minimal communication technology switching and minimal delay. This multi-objective optimization can be formulated as follows:
\begin{align}
    \text{(P1):} \quad & \underset{x_{i,j,m,k}, \delta_{i,j,k}, \beta_i}{\textit{minimize}}
    \alpha_1 \cdot \frac{\sum_{i,j,k,m} \left( x_{i,j,m,k} \cdot SE^m \right)}{S_1} + \notag \\
    &  \alpha_2 \cdot \frac{\sum_{i} \sum_{k} \left( s_{i,k}-s_{i,k-1} \right) ^2}{S_2} + \alpha_3 \cdot 
    \frac{\sum_{i,j,k} \delta_{i,j,k}}{S_3}, \label{P1} \notag \\ 
    & \hspace{-0.8cm}\textit{Subject to: } \notag 
\eqref{const2}, \eqref{const3}, \eqref{const4}, \eqref{const5}, 
    \eqref{const6}, \eqref{const7}, \eqref{const8}, \eqref{const9},  \eqref{const10}, \text{and } \eqref{const11}. \notag
\end{align}

In (P1), the terms $S_1$, $S_2$, and $S_3$ are normalization coefficients that take the maximum possible value for each multi-objective weight. However, such normalization alone often proves insufficient, especially in the case where the sub-objectives are \textit{conflicting}. Furthermore,  certain terms can disproportionately dominate the optimization process or cause solution saturation. This issue arises because each sub-objective function has a different sensitivity to the input change, causing some to inherently carry more weight even when normalized. Therefore, to mitigate the scaling issue between the terms, we introduce the regularization factors $\alpha_1$, $\alpha_2$, and $\alpha_3$, where $\sum_{i=1}^{3} \alpha_i =1$. The term $SE^m$ represents the energy needed to perform complete transmission using technology $m$. The optimization problem (P1) is indeed convex and classified as a Mixed Integer Non-Linear Program (MINLP).

\section{Simulation Results}

\begin{figure}[t]
\begin{center}\hspace{-0.3cm}\includegraphics[width=9cm]
{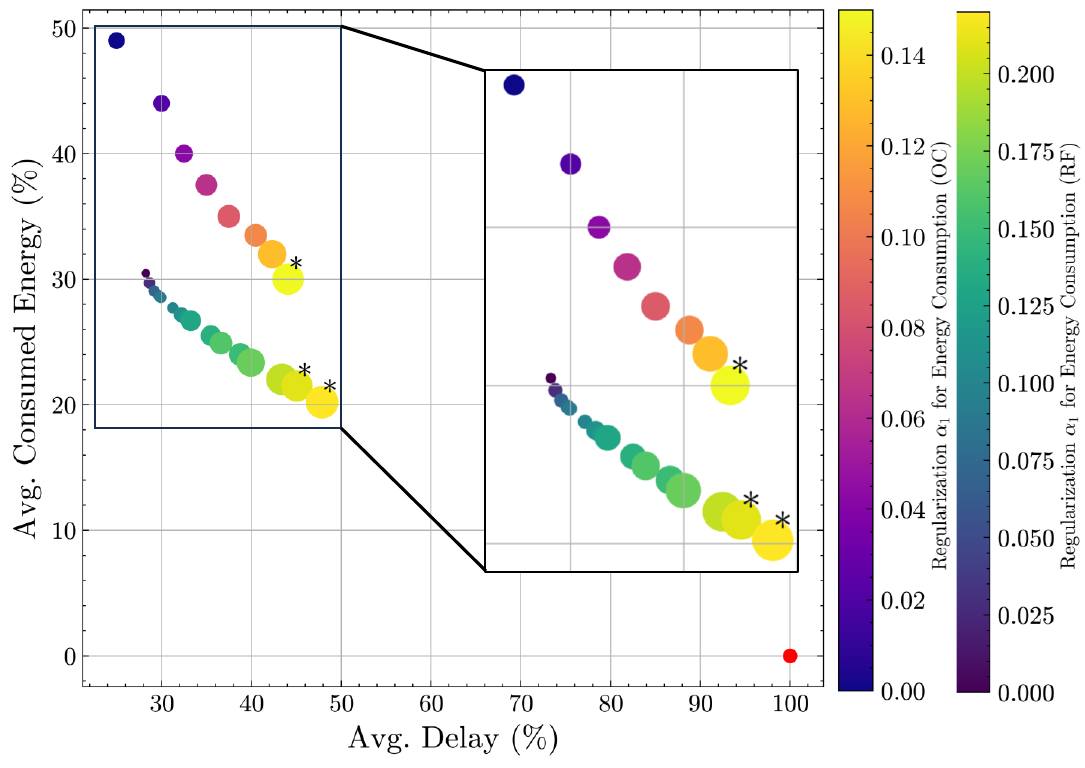}\end{center}
\caption{Avg. consumed energy (\%) vs Avg. Delay (\%), For RF and OC systems. The figure shows the feasible regularization $\alpha_1$ for each standalone system. The star represents the values of $\alpha_1$ with a delay of zero for the assigned communication. }\label{fig4}

\end{figure}
\subsection{Simulation Setup and AoI Metrics}

To simulate the behavior of a hybrid RF-OC IoT Network, we randomly generate the visibility matrix $V_m, m \in M$ to emulate the Packet Delivery Rate (PDR).  The visibility matrix follows a truncated continuous normal distribution centered at $0.85$ and $0.9$, with a maximum value of 1 for RF and OC, respectively. We set the threshold $\sigma_{m}$ for both communication technologies to be at least $97\%$ to consider only links with extremely high reliability. We consider a time frame of $T=200 ms$ with a time step of $10 ms$ (i.e., a total of 20 time steps). The variable ${E_s}^0$ for RF is set to $70mW$ while  ${E_s}^1$ for OC is set to $100mW$.   ${E_r}^0$ for RF is $10mW$ while  ${E_r}^1$ for OC is $7mW$. The total available energy of all nodes is randomly generated with a uniform distribution between $500$ and $700$ for both RF and OC technologies. The communication matrix $P$ is  randomly generated to contain zeros and ones to indicate if there is a transmission between a pair of devices at any given time. Similarly, the number of messages for each pair is also generated between $1$ and $5$ with a maximum length of $4$ time steps. The number of available types of data is set to $L=2$.

In our experiments, all the simulations are achieved using Monte-Carlo with $10,000$ iterations. All algorithms are implemented in a Python 3.7 environment and run on a 72-socket Intel(R) Xeon(R) Gold 5220 CPU @ 2.20GHz  with 256GB of RAM. For solving the MIQP problem, we use the academic CPLEX, an off-the-shelf optimization software package. For faster results, we set the duality gap in CPLEX to be 2\%.  We consider M-AoI and P-AoI to be the primary metrics for AoI analysis.  Given updates are received at times \( k_1, k_2, \ldots, k_n \) over a period \( T \), with \( \Delta(k_i^-) \) denoting the age just before the \( i \)-th update, the mean AoI and peak AoI can be computed as follows:
\begin{align}
\text{M-AoI} = \frac{1}{T} \int_0^T \Delta(k) \, dk \\
\text{P-AoI} = \frac{1}{N} \sum_{i=1}^N \Delta(k_i^-)
\end{align}
M-AoI provides an average measure of the AoI across the network, indicating overall system performance, while P-AoI captures the maximum age that any piece of information reaches before being updated. The P-AoI offers insight into the worst-case scenarios of data freshness.

\begin{table}[t]
\centering
\caption{M-AoI and P-AoI for both RF and RF-OC configurations computed for the overall system, data type 1, and data type 2.}
\label{AoItable}
\resizebox{8.7cm}{!}{\begin{tabular}{c|c|c|}
\cline{2-3}
                                    & RF                                                             & RF-OC                                                        \\ \hline
\multicolumn{1}{|c|}{System}        & \begin{tabular}[c]{@{}c@{}}M-AoI: 33.7, P-AoI: 40\end{tabular} & \begin{tabular}[c]{@{}c@{}}M-AoI: 16.2, P-AoI: 26\end{tabular} \\ \hline
\multicolumn{1}{|c|}{Data Type 1} & \begin{tabular}[c]{@{}c@{}}M-AoI: 18.3, P-AoI: 44\end{tabular} & \begin{tabular}[c]{@{}c@{}}M-AoI: 14.8, P-AoI: 33\end{tabular} \\ \hline
\multicolumn{1}{|c|}{Data Type 2} & \begin{tabular}[c]{@{}c@{}}M-AoI: 21.2, P-AoI: 38\end{tabular} & \begin{tabular}[c]{@{}c@{}}M-AoI: 17.4, P-AoI: 32\end{tabular} \\ \hline
\end{tabular}}
\end{table}

\begin{figure}[t]
 \vspace{0.4cm}
 \subfloat[]{%
      \includegraphics[width=0.245\textwidth]{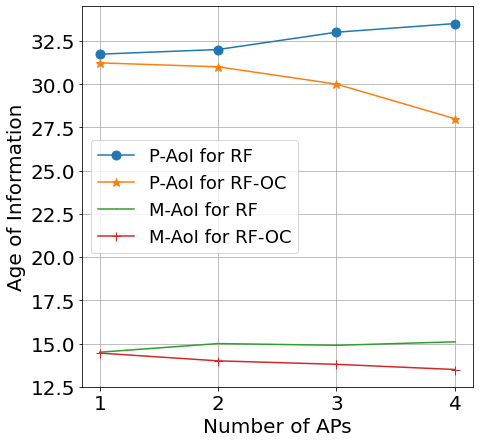}}
      \label{fig15}
 \qquad
  \hspace{-0.8cm}
 \subfloat[]{%
      \includegraphics[width=0.235\textwidth]{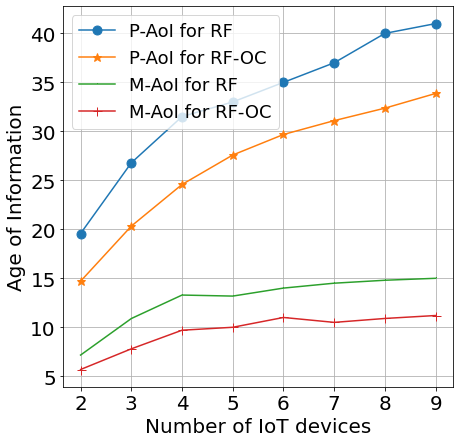}}
      \label{fig16}
       \caption{ AoI vs. Number of APs (a) and  AoI vs. Number of IoT nodes (b) for RF and Hybrid RF-OC systems.}%
        \label{fig:extra}
\end{figure}

\subsection{Sensitivity Analysis and Regularization}

In this section, we study the sensitivity of the sub-objective weights toward the optimization problem and communication selection.

In the first simulation, we study two different systems: 1) the first system uses RF technology and 2) The second system uses only OC technology. For both these systems, we consider $N_d=8$ with $N_{APs}=5$. In this simulation setting and since we are considering a single technology system, the second term in (P1) is eliminated and the optimization problem is reduced to a bi-objective problem with two conflicting sub-objective functions. In Fig.~\ref{fig4}, we evaluate the remaining two sub-objectives, mainly the avg. delay (\%) vs. the avg. consumed energy (\%), for different values of regularization term $\alpha_1$. The first observation is a Phillips curve for both systems which confirms the conflicting sub-objectives.

For the RF system, we can see a high sensitivity towards the regularization term $\alpha_1$. This is shown by the fact that, when increasing the value of $\alpha_1$, we see a jumps-like behavior with a decreasing avg. consumed energy (\%)  and an increasing avg. delay (\%) that starts in short intermittent levels and ends with a rapid movement. This can be explained by the fact that all the nodes have the same communication energy cost.  As $\alpha_1$ increases, minimizing energy becomes progressively more important compared to minimizing delay. Beyond a $\alpha_1=~0.175$, the energy savings from dropping multiple nodes outweigh the delay penalty, leading to a sudden reduction in selected nodes. The mathematical formulation of the objective function and constraints have certain non-linearities and interactions between variables that are not immediately apparent but manifest in the solutions as abrupt changes. This behavior where the optimizer starts dropping all nodes at once rather than one by one suggests that the marginal benefit (in terms of the objective function) of removing any single node does not make a significant difference until a certain threshold in weights is reached, hence the jumps. For the OC system, we notice the same observation with different acceptable values of $\alpha_1$. The key difference for the OC system is that the jumps between the solutions start at earlier values of $\alpha_1$ compared to the OC system. Also, the number of points (i.e., solution levels) is fewer due to the OC being more aggressive in eliminating the communication of IoT devices as they are more expensive in the OC setting. 
\begin{figure}[t]
\centering
 \subfloat[]{%
      \includegraphics[width=0.33\textwidth]{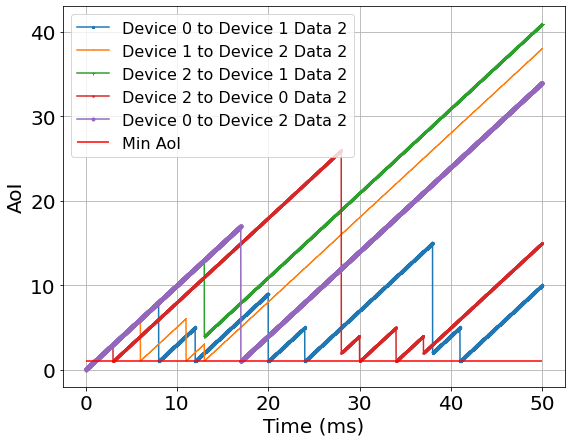}}
      \label{fig1}
 \qquad
 \hspace{-0.8cm}
 \centering
 \subfloat []{%
      \includegraphics[width=0.33\textwidth]{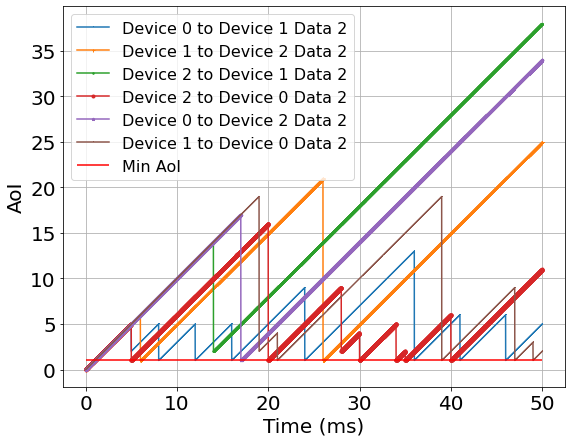}}
      \label{fig2}
       \caption{AoI for RF (a) and AoI for RF-OC (b) vs. time.}%
        \label{fig:images5}
\end{figure}

For both settings, we can see that with value $\alpha_1=0$, we get the highest avg. consumed energy (\%) with lowest avg. delay (\%) as this is due to the second term in (P1) dominating the optimization problem. The red dot represents the avg. consumed energy (\%) and  avg. delay (\%) for all values of $\alpha_1>~0.23$ for RF systems, and for values of $\alpha_1>~0.16$ for OC systems. This is because above these two values, the first term in (P1) for each system starts dominating and the optimal solution becomes an empty set. The dots marked with $^*$ show the different configurations where the selected nodes for communication have a delay of $0$. This confirms that the result of increasing $\alpha_1$ leaves at some point devices that can only communicate as soon as their messages are generated. The optimizer starts dropping devices with a higher delay, for both OC and RF, leaving only the ones with the lowest delay. Indeed, Fig.~\ref{fig4} also shows the different regularization terms to pick from if we aim to guarantee a certain minimum of energy consumption or delay, for example. Note that the value of $\alpha_2$ in (P1) for the RF-OC system can be selected by: 1) selecting a value of $\alpha_1$ from the previous analysis that falls within, both RF and OC, acceptable regions and 2) performing a grid search over the possible values of $\alpha_2 \in [0,1]$ and selecting those that makes the individual values of the sub-objective function in (P1) no larger than a certain defined threshold, say $5\%$,  from their values when optimized separately. In the following simulations, we set $\alpha_1=0.1$, $\alpha_2=0.1$, and consequently, $\alpha_3=0.8$.

\begin{figure}[t]
 \subfloat[]{%
      \includegraphics[width=0.238\textwidth]{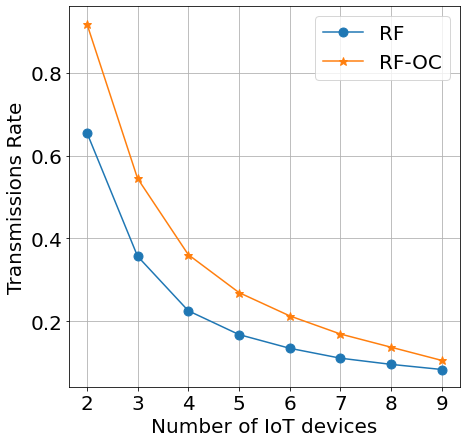}}
      \label{fig1}
 \qquad
 \hspace{-0.8cm}
 \subfloat []{%
      \includegraphics[width=0.245\textwidth]{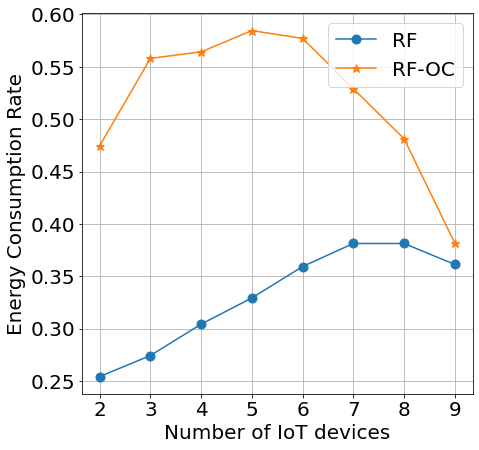}}
      \label{fig2}
       \caption{Avg. Trans. Rate (a) and Avg. Consumed Energy (b) vs. Number of IoT devices for RF and RF-OC with $N_{APs}=3$.}%
        \label{fig:images1}
\end{figure}

\subsection{Simulation Results}

In the second simulation, we perform a comparison between a network that has only RF capabilities and a hybrid network with OC and RF capabilities.  In Table.~\ref{AoItable},  we consider $N_d=9$, $N_{APs}=2$, and present comparative results for M-AoI and P-AoI across these two configurations. For the overall system with only one type of data, the RF-OC configuration demonstrates significantly better performance, with a M-AoI of $16.2$ compared to $33.7$ in the RF configuration, and a P-AoI of 26 versus 40 in RF. This indicates a substantial improvement in the freshness of information transmitted via the hybrid system. We see that for a system with two data types, the improvement with RF-OC remains consistent though less pronounced, with lower M-AoI and P-AoI for RF-OC. This indeed confirms that not only does the overall system have lower M-AoI and P-AoI, but also each type of data has a reduced information age, which can be crucial for applications requiring rapid data updates.  In Fig.~\ref{fig:extra}, We study the P-AoI and M-AoI on RF and RF-OC systems. We see that both metrics increase with the number of devices, indicating that the information becomes less current as more devices are added. The two AoI metrics are lower for the RF-OC system than the RF system with about $15\%$. We also see that the M-AoI is more stable with less variation than the P-AoI for both networks. When adding new APs,  the M-AoI and P-AoI increase in the RF system while they decrease with new APs in the RF-OC system. This is mainly because APs are enabled by OC, and hence, in the RF-OC system, they alleviate the communication between the IoT nodes and rely on the OC medium. In Fig.~\ref{fig:images5}, we show the time evolution of AoI for RF and RF-OC for type-2 data. We can see that  RF-OC has more frequent updates with lower overall AoI. Also, the number of communicating devices is higher than those of the RF system.

\begin{figure}[t]
 \vspace{0.05cm}
 \subfloat[]{%
      \includegraphics[width=0.245\textwidth]{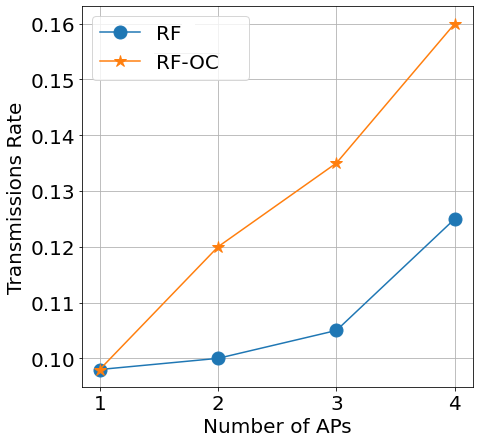}}
      \label{fig1}
 \qquad
 \hspace{-0.9cm}
 \subfloat[]{%
      \includegraphics[width=0.2445\textwidth]{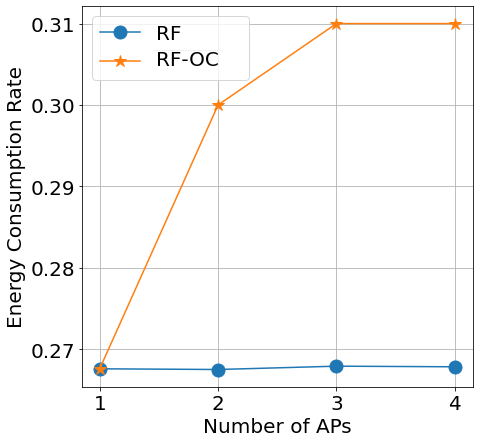}}
      \label{fig2}
       \caption{ Avg. Transmission Rate  (a) and Avg. Energy Consumption Rate (b) vs. Number of APs for RF and Hybrid RF-OC systems.}%
        \label{fig:images3}
\end{figure}

In Fig.~\ref{fig:images1}, we consider a fixed number of APs with $N_{APs}=3$ and vary $N_{d}$. We see that both RF and RF-OC systems illustrate a decreasing trend in transmission rate as the number of IoT nodes increases. RF-OC starts higher but decreases more sharply than RF system. This trend suggests that as more nodes are added to the network, the network becomes more congested, leading to a decrease in the efficiency of data transmission. The hybrid RF-OC system seems to manage the increase in devices slightly better, potentially due to the additional bandwidth. The overall average energy consumption increases with the increase in the number of IoT nodes for both RF and RF-OC, with RF-OC having a higher rate of increase compared to RF. The increase is generally due to the introduction of new nodes in the network that will take part in the communication. The RF-OC system has more opportunities for communication with the increase of nodes. At some point, the energy consumption starts to fall back, and this can be explained by the congestion of the network.

In Fig.~\ref{fig:images3}, we consider a fixed number of IoT nodes with $N_{d}=10$ and vary, $N_{APs}$, the number of IoT APs. We notice that the RF-OC network achieves a higher transmission rate of $~25\%$ and energy consumption of $10\%$ more than the RF network. The higher energy consumption of RF-OC is due to OC communication consuming more energy per transmission than RF communication.

\section{Conclusion and Future Work}

This paper presents an optimization algorithm for IoT networks that utilizes hybrid nodes combining OC and RF. Our approach dynamically selects the most appropriate communication technology based on multi-objective optimization, significantly reducing the AoI and enhancing data freshness. Despite its effectiveness, the algorithm faces challenges in scalability and adaptability due to the NP-hard nature of the problem and the dynamic IoT environments. Future enhancements will focus on integrating adaptive decision-making to accommodate real-time environmental changes with varying time windows and incorporating smoothing constraints to ensure more stable transitions between communication technologies. Also, future directions will involve AI-driven approaches to enable fast communication scheduling in large-scale IoT networks


\section*{Acknowledgment}

This research was supported by the SUPERIOT project. The SUPERIOT project has received funding from the Smart Networks and Services Joint Undertaking (SNS JU) under the European Union’s Horizon Europe research and innovation program under Grant Agreement No. 101096021. The work of Hazem Sallouha was funded by the Research Foundation Flanders (FWO), Postdoctoral Fellowship No. 12ZE222N.




\bibliographystyle{IEEEtran}
\bibliography{ref}

\begin{thebibliography}{10}
\providecommand{\url}[1]{#1}
\csname url@samestyle\endcsname
\providecommand{\newblock}{\relax}
\providecommand{\bibinfo}[2]{#2}
\providecommand{\BIBentrySTDinterwordspacing}{\spaceskip=0pt\relax}
\providecommand{\BIBentryALTinterwordstretchfactor}{4}
\providecommand{\BIBentryALTinterwordspacing}{\spaceskip=\fontdimen2\font plus
\BIBentryALTinterwordstretchfactor\fontdimen3\font minus \fontdimen4\font\relax}
\providecommand{\BIBforeignlanguage}[2]{{%
\expandafter\ifx\csname l@#1\endcsname\relax
\typeout{** WARNING: IEEEtran.bst: No hyphenation pattern has been}%
\typeout{** loaded for the language `#1'. Using the pattern for}%
\typeout{** the default language instead.}%
\else
\language=\csname l@#1\endcsname
\fi
#2}}
\providecommand{\BIBdecl}{\relax}
\BIBdecl

\bibitem{10433153}
G.~Rathee, C.~A. Kerrache, C.~T. Calafate, M.~Bilal, and H.~Song, ``Smart: A secure remote sensing solution for smart cities’ urban areas,'' \emph{IEEE Sensors Journal}, vol.~24, no.~7, pp. 11\,553--11\,561, 2024.

\bibitem{10032496}
Y.~Chen, J.~Wang, X.~Wang, and J.~Song, ``Age of information optimization in multi-channel network with sided information,'' \emph{IEEE Communications Letters}, vol.~27, no.~3, pp. 1030--1034, 2023.

\bibitem{10000990}
D.~e.~a. Deng, ``Information freshness in a dual monitoring system,'' in \emph{2022 IEEE Global Communications Conference}, 2022, pp. 4977--4982.

\bibitem{10049773}
B.~Yu \emph{et~al.}, ``Adaptive packet length adjustment for minimizing age of information over fading channels,'' \emph{IEEE Transactions on Wireless Communications}, vol.~22, no.~10, pp. 6641--6653, 2023.

\bibitem{katz2024towards}
M.~Katz, T.~Paso, K.~Mikhaylov, L.~Pessoa, H.~Fontes, L.~Hakola, J.~Lepp{\"a}niemi, E.~Carlos, G.~Dolmans, J.~Rufo \emph{et~al.}, ``Towards truly sustainable iot systems: the superiot project,'' \emph{Journal of Physics: Photonics}, vol.~6, no.~1, p. 011001, 2024.

\bibitem{2}
H.~Abuella, M.~Elamassie, M.~Uysal, Z.~Xu, E.~Serpedin, K.~A. Qaraqe, and S.~Ekin, ``Hybrid rf/vlc systems: A comprehensive survey on network topologies, performance analyses, applications, and future directions,'' \emph{IEEE Access}, vol.~9, pp. 160\,402--160\,436, 2021.

\bibitem{10008721}
A.~F. Raouf, C.~K. Anjinappa, and G.~I., ``Optimal design of energy-harvesting hybrid vlc/rf networks,'' in \emph{2022 IEEE Globecom Workshops (GC Wkshps)}, 2022, pp. 705--710.

\bibitem{9912428}
A.~Hamrouni, A.~Khanfor, H.~Ghazzai, and Y.~Massoud, ``Context-aware service discovery: Graph techniques for iot network learning and socially connected objects,'' \emph{IEEE Access}, vol.~10, pp. 107\,330--107\,345, 2022.

\bibitem{9380899}
R.~D. Yates, Y.~Sun, D.~R. Brown, S.~K. Kaul, E.~Modiano, and S.~Ulukus, ``Age of information: An introduction and survey,'' \emph{IEEE Journal on Selected Areas in Communications}, vol.~39, no.~5, pp. 1183--1210, 2021.

\bibitem{8883204}
A.~Rovira-Sugranes and A.~Razi, ``Optimizing the age of information for blockchain technology with applications to iot sensors,'' \emph{IEEE Communications Letters}, vol.~24, no.~1, pp. 183--187, 2020.

\bibitem{1}
M.~Fakirah, S.~Leng, M.~Mohammad, A.~Abualnor, and M.~L. Betalo, ``Improving user data rate performance for hybrid rf/vlc outdoor vehicular communications,'' in \emph{2022 IEEE 22nd International Conference on Communication Technology (ICCT)}, 2022, pp. 812--816.

\bibitem{3}
M.~Obeed \emph{et~al.}, ``Joint optimization of power allocation and load balancing for hybrid vlc/rf networks,'' \emph{Journal of Optical Communications and Networking}, vol.~10, no.~5, pp. 553--562, 2018.

\bibitem{5}
Y.~Xiao, P.~D. Diamantoulakis, Z.~Fang, L.~Hao, Z.~Ma, and G.~K. Karagiannidis, ``Cooperative hybrid vlc/rf systems with slipt,'' \emph{IEEE Transactions on Communications}, vol.~69, no.~4, pp. 2532--2545, 2021.

\bibitem{9755889}
Y.~Liu, Z.~Chang, G.~Min, and S.~Mao, ``Average age of information in wireless powered mobile edge computing system,'' \emph{IEEE Wireless Communications Letters}, vol.~11, no.~8, pp. 1585--1589, 2022.

\bibitem{9239322}
J.~Xu and N.~Gautam, ``Peak age of information in priority queuing systems,'' \emph{IEEE Trans. on Information Theory}, vol.~67, no.~1, 2021.

\bibitem{articless}
M.~Cococcioni and L.~Fiaschi, ``The big-m method with the numerical infinite m,'' \emph{Optimization Letters}, vol.~15, 10 2021.

\end{thebibliography}
\end{document}